\documentclass[aps,superscriptaddress,nofootinbib]{revtex4}

\newcommand{\beq}{\begin{equation}}
\newcommand{\eeq}{\end{equation}}
\newcommand{\bea}{\begin{eqnarray}}
\newcommand{\eea}{\end{eqnarray}}
\newcommand{\ce}{{\mathcal{E}}}
\newcommand{\cj}{{\mathcal{J}}}

\newcommand{\cQ}{{\mathcal{Q}}}

\newcommand{\cD}{{\mathcal{D}}}

\newcommand{\cW}{{\mathcal{W}}}
\newcommand{\cV}{{\mathcal{V}}}
\newcommand{\te}{{\tilde{\epsilon}}}
\newcommand{\tet}{{\tilde{\eta}}}
\newcommand{\grv}{{4\pi\over m_{\rm pl}^2}}
\newcommand{\invgrv}{{m_{\rm pl}^2\over 4\pi}}

\newcommand{\dk}{{\int{d^3k\over (2\pi)^{3\over 2}}}}

\newcommand{\e}{{e^{i\mathbf{k}\mathbf{x}}}}

\begin{document}

\title{Non-linear inflationary perturbations}

\author{G.I. Rigopoulos}
\affiliation{Spinoza Institute, University of Utrecht\\
 Postbus 80.195, 3508 TD Utrecht\\ The Netherlands}

\author{E.P.S. Shellard}
\affiliation{ Department of Applied Mathematics and Theoretical Physics\\
Centre for Mathematical Sciences\\
University of Cambridge\\
Wilberforce Road, Cambridge, CB3 0WA, UK}

\begin{abstract}
\noindent We present a method by which cosmological perturbations
can be quantitatively studied in single and multi-field
inflationary models beyond linear perturbation theory. A
non-linear generalization of the gauge-invariant Sasaki-Mukhanov
variables is used in a long-wavelength approximation. These
generalized variables remain invariant under time slicing changes
on long wavelengths. The equations they obey are relatively simple
and can be formulated for a number of time slicing choices.
Initial conditions are set after horizon crossing and the
subsequent evolution is fully non-linear. We briefly discuss how
these methods can be implemented numerically to the study of
non-Gaussian signatures from specific inflationary models.

\end{abstract}

\maketitle

\section{Introduction}

Our current understanding of the universe on large scales
inevitably leads to questions regarding its earliest moments. The
patterns seen in the distribution of matter on such scales and the
fluctuations in the temperature of the CMB, point to mechanisms in
the early universe that could have imprinted the initial
conditions for such patterns on the otherwise homogeneous
primordial universe. Recent data on CMB anisotropies \cite{wmap}
leave no doubt that initial fluctuations existed on scales much
larger than the causal horizon at the time of recombination.
Inflation \cite{guth} is a concept which can explain the origin of
such fluctuations \cite{mc, inflation} as well as many other
special features of the cosmos.

Even though the inflationary paradigm has been developing for over
twenty years now, there is currently no agreed upon model of the
inflationary epoch. Almost all models of inflation invoke one or
more scalar fields to drive an initial phase of accelerated
expansion. The latter is capable of producing fluctuations in the
energy density by amplifying the quantum fluctuations of any light
scalar field present during inflation. These fluctuations are
usually represented as linearized deviations from a homogeneous
evolution and are therefore described by non-interacting quantum
fields living in the expanding spacetime. Linearity, along with
the additional assumption that the initial state is the vacuum (as
defined on scales much smaller then the Hubble radius), lead to
the prediction that inflation creates Gaussian
fluctuations.\footnote{In contrast, topological defects cannot be
described as a smooth linear deviation from a homogeneous
background and hence are intrinsically non-Gaussian.} The
smallness of the observed CMB anisotropy certainly justifies the
use of linear perturbation theory as a first approximation.
However, some non-linearity will always be present in inflation
due to the non-linear nature of gravity and the fact that the
potential responsible for inflation may be interacting. Another
property that is usually attributed to inflation is that it
produces adiabatic fluctuations. This is strictly true only for
single-field models. When more scalar fields are present, there is
also a possibility for isocurvature perturbations. It has been
suggested in the past that the interplay of isocurvature
perturbations and non-linearity can lead to enhancement of the
non-Gaussianity of primordial fluctuations. Reference \cite{uzan}
is an example of this; the curvaton paradigm \cite{curvaton}
provides another possibility \cite{ld}.

In this paper we present a method for studying the evolution of
perturbations in an arbitrary scalar-field-driven inflationary
model. The equations are valid in the long wavelength regime
(scales larger than the Hubble radius) and allow for the
calculation of non-linear effects. Scalar metric perturbations are
taken into account and the scheme can in principle be formulated
with an arbitrary time slicing. However, we find the use of a
particular choice of time more convenient. The main variables of
this formalism are non-perturbative generalizations of the useful
Sasaki-Mukhanov variables of linear theory and are invariant under
time slicing changes on long wavelengths; one can say that they
are gauge invariant beyond perturbation theory. Initial conditions
are provided after horizon crossing from linear perturbation
theory. We argue that this is enough and any observable
non-linearity will be produced by the subsequent evolution.

The outline of the paper is as follows: In section 2 we provide a
framework for the analysis of the long-wavelength dynamics for a
general inhomogeneous inflationary model. The approach goes beyond
perturbation theory and allows for a treatment of all possible
non-linearities. It is technically much simpler than higher order
perturbation approaches \cite{secondorder}. In section 3 we
supplement the long-wavelength equations by stochastic terms which
set the initial conditions after horizon crossing. Although
formulated for a general time slicing, we find that a particular
choice of time is most convenient. In section 4 we discuss
progress towards a numerical implementation of these methods and
the future prospects.

\section{Long wavelength dynamics}

A key characteristic of the inflationary era is the behaviour of
the comoving Hubble radius $\left(aH\right)^{-1}$ which shrinks
quasi-exponentially. This behaviour is unlike what happens during
the radiation and matter eras where this scale grows. This feature
allows inflation to answer a number of puzzling facts about the
universe and also provides a mechanism for the quantum generation
of fluctuations \cite{mc, inflation}. All of the astrophysically
relevant scales start their lives subhorizon but eventually they
are stretched to superhorizon sizes. In this paper we will focus
on the superhorizon regime. Hence, a long-wavelength approximation
along with a way to set initial conditions suffices for our
purposes. The long-wavelength approximation we employ consists of
dropping from the equations all terms containing second order
spatial gradients \cite{sb1, long
  wavelength, GP}. Then, spacetime can be described by the metric
\footnote{We restrict
  attention to metrics with zero shift $N_i=0$}
\beq
ds^2=-N^2(t,\mathbf{x})dt^2+a^2(t,\mathbf{x})h_{ij}(\mathbf{x})dx^idx^j,
\eeq where $N$ is the lapse function, $a$ is a local scale factor
and $h_{ij}$ is a local spatial metric with unit determinant. In
linear perturbation language, the latter contains one scalar
degree of freedom (d.o.f), which enters as a second order
  spatial gradient and hence
can be ignored, vector and tensor d.o.f's. It can be shown quite
generally that $h_{ij}$ freezes on superhorizon scales \cite{sb1}.
The local scale factor $a$ contains the other scalar d.o.f that
carries the dynamics on these scales. $N$ can be chosen freely and
corresponds to a choice of time slicing.

It turns out that under the above long-wavelength approximation
the equations of motion are reduced to those of a homogeneous
Friedmann-Lemaitre-Robertson-Walker (FLRW) cosmology, applied
locally, plus a gradient constraint linking spatial gradients in
matter and geometry. Hence, each point evolves as an independent
universe with its own value for the matter fields, Hubble rate and
scale factor, provided initial conditions which satisfy the
gradient constraint have been specified \cite{sb1, GP}. In
particular, after dropping a decaying mode \cite{sb1,GP}, the
evolution equations become \bea
\frac{dH}{dt}&=&-\grv N\left(\ce+{1 \over 3}S\right)\,, \\
\frac{d\ce}{dt}&=&-3NH\left(\ce+{1 \over 3}S\right)\,,  \eea and
are supplemented by two constraints \beq H^2=\frac{8\pi}{3m_{\rm
pl}^2}\ce\,,\quad
\partial_iH=\frac{4\pi}{m_{\rm pl}^2}\cj_{i}.
\eeq Here, matter is described by the energy momentum tensor
$T_{\mu\nu}$. We have defined the energy density, momentum density
and stress tensor \beq \ce\,{\equiv}N^{-2}T_{00}\,,\quad
\cj_{i}\equiv -N^{-1}T_{0i}\,,\quad S_{ij}\equiv{}T_{ij}\,, \eeq
and the local expansion rate \beq\label{ler} H\equiv{1\over
N}{d\over dt}\left(\ln{a}\right)\,. \eeq A vertical bar denotes a
covariant derivative with respect to the spatial metric. The above
system is consistent only if $S^i{}_j=\frac{1}{3}S\delta^i_j$
which is expected to be true on long wavelengths.

We now focus on an inflationary era driven by $n$ scalar fields
with the energy momentum tensor \beq
T_{\mu\nu}=G_{AB}\partial_{\mu}\phi^A\partial_{\nu}\phi^B-g_{\mu\nu}\left(\frac{1}{2}G_{AB}\partial^{\lambda}\phi^A\partial_{\lambda}\phi^B+V\right),
\eeq so that \bea
\ce&\simeq&\frac{1}{2}G_{AB}\Pi^A\Pi^B+V(\phi)\,,\\
\cj_i&=&-G_{AB}\Pi^A\partial_i\phi^B\,,\\
S_{ij}&\simeq&a^2
h_{ij}\left(\frac{1}{2}G_{AB}\Pi^A\Pi^B-V\right), \eea where the
approximate equality indicates second order spatial gradients
dropped. We defined the field momentum as \beq
\Pi^A={\dot\phi^A\over N}\,. \eeq Then, the long-wavelength
equations of motion for this system are \bea
\label{H_dynamics}\frac{dH}{dt}&=&-\frac{4\pi}{m_{\rm pl}^2}N\Pi_B\Pi^B\,, \\
\label{momentum_dynamics}
{\cD_t}\Pi^A&=&-3NH\Pi^A-NG^{AB}V_B\,,\\
\label{friedmann}H^2&=&\frac{8\pi}{3m_{\rm pl}^2}\left(\frac{1}{2}\Pi_B\Pi^B+V\right)\,,\\
\label{0i} \partial_iH&=&-\frac{4\pi}{m_{\rm
pl}^2}\Pi_B\partial_i\phi^B, \eea where $V_B\equiv\partial_BV$ and
the symbol $\cD_t$ appearing in (\ref{momentum_dynamics}) will be
defined shortly. Note that we have taken a general field metric
$G_{AB}$ and all the equations below will be valid for such an
arbitrary metric. We thus view the dynamics as taking place on a
general n-dimensional field manifold where $\phi^A$ are just a set
of $n$ functions parameterizing it. Hence, it makes sence to
define covariant derivatives when considering the spatial or
temporal dependence of various quantities. For a spacetime
dependent quantity $L^A(t,\mathbf{x})$ which transforms as a
vector in field space, we define the covariant derivatives
\beq\label{cov_1} \cD_t{L^A}= \partial_tL^A
+\Gamma^A_{BC}\,\partial_t\phi^BL^C\,,\quad \cD_i{L^A}=
\partial_iL^A +\Gamma^A_{BC}\,\partial_i\phi^BL^C\,, \eeq with
$\Gamma^A_{BC}$ the symmetric connection formed from $G_{AB}$. The
quantities $\partial_i\phi^B$ and $N\Pi^B$ transform as vectors in
field space but $\phi^B$ does not.

As mentioned above, the system of eqs (\ref{H_dynamics}) -
(\ref{friedmann}) is exactly the same as the equations of FLRW
cosmology applied locally. However, equation (\ref{0i}) gives an
additional constraint which is obviously absent in the homogeneous
case. If one drops the $\Pi^2$ term from the Friedmann equation
(\ref{friedmann}) and the $\cD_t\Pi^A$ term from eq
(\ref{momentum_dynamics}), eqs (\ref{friedmann}) and (\ref{0i})
are equivalent. This assumption is implicit in the "separate
universe" picture of non-linear evolution which is a simple way to
view the long wavelength dynamics (see eg. \cite{ld} and
references therein). However, in the case of several fields
certain components of $\cD_t\Pi^A$ may be important even though
slow-roll inflation is still valid \cite{bartjan}, and ignoring
them may miss the relevant effects. No such assumption needs to be
made for the formalism we develop here.

A natural way to parameterize inhomogeneity beyond perturbation
theory is to use spatial gradients of various quantities of
interest. This is similar in spirit to the approach first
advocated in \cite{eb}. In general, the value of any spatial
gradient will depend on the chosen time-slicing and by an
appropriate choice of time slices the inhomogenous part of any
spacetime scalar can be made to vanish. However, one can form
combinations of spatial gradients, including the gradient of the
integrated expansion (the local scale factor), which are invariant
under long wavelength changes of time slicing. They can be
constructed similarly to those of linear perturbation theory. One
such variable is
 \beq\label{g.i.var2}
\cQ_i^A=a\left(\partial_i\phi^A-{\Pi^A \over H}X_i\right)\,, \eeq
with $X_i \equiv \partial_i \ln a$. Such combinations were first
introduced in \cite{GP} and it can be checked explicitly that they
are invariant under $t\rightarrow T(t,\mathbf{x})$ transformations
on long wavelengths. More recently, such variables were considered
by the authors of \cite{langlois_vernizzi} in a more geometric
framework who also showed explicitly that they reproduce
previously introduced gauge-invariant perturbations at second
order. Note that when linearized around a homogeneous background,
$\cQ_i^A$ is just the gradient of the well known Sasaki-Mukhanov
variable \cite{mfb} \beq \label{s-m} \delta
q^A=a\left(\delta\phi^A-{\dot{\phi}^A\over NH}\Psi\right), \eeq
where $\Psi=\delta a/a$ is the perturbation in the trace of the
spatial metric. In our notation, the linear version of $X_i$ is
just $\partial_i\left(\delta a\right)/ a(t)=\partial_i\Psi$. A
similar gauge-invariant variable is \cite{GP} \beq\label{g.i.var3}
\zeta_i=X_i-{ NH \over \dot{H} }\partial_i\ce\,\, \eeq which,
using eqs (\ref{0i}) and (\ref{g.i.var2}), ($\ref{g.i.var3}$) can
also be written as \beq \zeta_i=\grv{1 \over a}{ NH \over \dot{H}
} \Pi_A\cQ^A_i. \eeq This is a non-linear generalization of the
well known linear curvature variable $\zeta$.

The $\cQ_i^A$'s obey relatively simple equations of motion which
can be derived by taking spatial derivatives of equations
($\ref{H_dynamics}$) - ($\ref{friedmann}$) and using (\ref{0i}).
Details of the derivation can be found in Appendix A. It is
convenient to define a set of slow-roll parameters for the
multi-field model (see \cite{bartjan} for a discussion in the
framework of linear perturbation theory): \beq \te=-{\dot{H}\over
NH^2}=\grv{\Pi^2\over H^2}\,,\quad \tet^A={1\over
N}{\cD_t\Pi^A\over H\Pi}\,, \eeq which are usually assumed to be
much smaller than unity during inflation.\footnote{We note here
that the following equations, although expressed in terms of the
slow roll parameters, are exact, i.e no assumption has been made
about the smallness of the latter.} It will also be useful to
define \beq \omega^A=\sqrt{\te}{\Pi^A \over \Pi}. \eeq Then, the
following equations of motion can be derived for the $\cQ^A_i$'s
(see Appendix A) \beq\label{basic} {\cD_t^2
}\cQ^A_i-\left({\dot{N} \over
N}-NH\right){\cD_t}\cQ^A_i+\Omega^A{}_B\cQ^B_i=0 \eeq with the
``mass matrix'' \beq \label{omega} \Omega^A{}_B=N^2V^A{}_B-\invgrv
(NH)^2R^A{}_{FCB}\omega^F\omega^C
-\left(NH\right)^2\left[\left(2-\te\right)\delta^A{}_B+2\left(3+\te\right)\omega^A\omega_B+2\
 \sqrt{\te}\left(\tet_B\omega^A+\tet^A\omega_B\right)\right],
\eeq $V^A{}_B=\cD_BV^A$ and $R^A{}_{FCB}$ the curvature tensor of
the field manifold. These are formally the same equations as those
of linear perturbation theory for the Sasaki-Mukhanov variables
with the $k^2$ terms dropped \cite{bartjan}. Note that although
equation (\ref{basic}) appears to be linear in the $\cQ^A_i$'s, it
incorporates the full non-linear dynamics on large scales since
its coefficients are spatially varying functions. They depend
implicitly on the $\cQ^A_i$'s via a set of relations which express
various local quantities in terms of them. Hence, the $\cQ^A_i$'s
can be seen as master variables encoding all the d.o.f's of the
inhomogeneous system. Expressing $\partial_i\phi^A$ in terms of
$\cQ^A_i$, using ($\ref{0i}$) and noting that
$\cD_t(\partial_i\phi^A)=\cD_i\left(N\Pi^A\right)$, we get \bea
\label{constr_1}
\partial_i\left(\ln H \right) &=& -{1\over a}\left(\grv\right)^{1\over
  2} \omega_A\cQ^A_i-\te X_i\,,\\
\label{constr_2}\partial_i \phi^A&=&{1\over
a}\cQ^A_i+\left(\invgrv\right)^{1\over
  2}\omega^A X_i,
\eea and \beq\label{constr_3} \cD_i\Pi^A={1 \over
aN}{\cD_t}\cQ^A_i-{H \over
  a}\left[\delta^A{}_C+\omega^A\omega_C\right]\cQ^C_i
+H\left(\invgrv\right)^{1\over 2}\sqrt{\te}\tet^AX_i. \eeq The
right hand sides of Eqs (\ref{constr_1}) - (\ref{constr_3}) are
given in terms of gauge invariant variables, apart from the terms
involving $X_i$. The latter cannot be expressed via a similar
relation of the form $X_i=f_A\cQ^A_i$. For a general choice of
time, (\ref{constr_1}) can also be written as \beq\label{Xi}
\dot{X}_i=-NH\te X_i+H\partial_iN-\left(\grv\right)^{1\over 2}{NH
\over a}\omega_A\cQ^A_i\,. \eeq Note that $X_i$ is not considered
as an independent d.o.f. since it is gauge-dependent. All physical
d.o.f's describing inhomogeneity are encoded in the $\cQ^A_i$'s.
Hence, a natural choice is to have $X_i=0$ if, at all times,
$\cQ^A_i= 0$ and $\dot{\cQ}^A_i= 0$. Such a choice simply means
setting $X_i(t_{init})=0$ initially and making $N$ some function
of $H$, $\phi^A$, $\Pi^A$ and $a$. This ensures that
$\partial_iN=f_A\cQ^A_i+g_A\dot{\cQ}^A_i+hX_i$ with $f_A\,,g_A$
and $h$ arbitrary functions.

The above equations are obviously valid for any choice of N and,
as emphasized above, $Q^A_i$ is gauge invariant. During inflation
however, it is natural to use the logarithm of the local value of
the Hubble radius as the time variable \cite{sb2} \beq
\label{time} t=\ln (aH)\quad\Leftrightarrow\quad t=\int H(1-\te)
dT\,, \eeq with T being proper time. For such a gauge choice we
get \beq \label{lapse} N^{-1}={dt\over dT}=H(1-\te). \eeq Since
$\partial_it=0$ by definition on surfaces of constant time, we
have \beq \label{Xi2} X_i=-\partial_i(\ln H)\,, \eeq and
(\ref{Xi}) is automatically satisfied. In this gauge the
constraints simplify: \bea \label{constraints_1}
X_i&=&\left(\grv\right)^{1\over 2} {1\over a(1-\te)}\omega_A\cQ^A_i=-\partial_i(\ln H)\,,\\
\partial_i\phi^A&=&{1\over
  a}\left[\delta^A{}_B+{\omega^A\omega_B\over (1-\te)}\right]\cQ^B_i\,,
\eea and \beq\label{constraints_2} \cD_i\Pi^A={H \over
a}(1-\te){\cD_t}\cQ^A_i-{H \over
  a}\left[\delta^A{}_C+\omega^A\omega_C-\sqrt{\te}\tet^A{\omega_C
  \over (1-\te)}\right]\cQ^C_i.
\eeq Another advantage of (\ref{time}) will be discussed in the
next section. Equations ($\ref{constraints_1}$) -
($\ref{constraints_2}$) specify the gradients of the local
quantities that appear in the coefficients of ($\ref{basic}$) in
terms of the $\cQ^A_i$'s. Note that in this slicing, for
quasi-exponential expansion, $t$ is monotonic. The validity of
such a time variable extends only up to the end of inflation when
the comoving horizon starts growing again.

Before closing this section we would like to make one final
remark. By taking the time derivative of equation (\ref{0i}) we
can derive the following relation between $\cD_t{\cQ}^A_i$ and
${\cQ}^A_i$ \beq \label{f.o.t}
\Pi_A\cD_t{\cQ}^A_i=NH\left[\left(1+\te\right)\Pi_A+\Pi\tet_A\right]\cQ^A_i.
\eeq Since $\cD_tQ^A_i$ is related to $\cQ^A_i$, there is an
apparent reduction of order for projections of the perturbations
along the $\Pi^A$ direction in field space. In particular, for the
single-field case the full dynamics on long wavelengths can be
simply written as a first order equation \beq\label{f.o.t.single}
d_t{\cQ}_i=NH\left(1+\te+\tet\right)\cQ_i\,. \eeq From
($\ref{f.o.t.single}$) we can obtain a non-linear conservation law
\cite{GP} (see also \cite{langlois_vernizzi}) in the single field
case. For \beq \zeta_i\equiv - \left(\grv\right)^{1\over 2}
{1\over a\sqrt{\te}}\cQ_i\,, \eeq  given that \beq
\dot{\te}=2NH\te\left(\te+\tet^B{\Pi_B\over \Pi}\right)\,, \eeq it
straightforward to calculate that $\zeta_i$ is conserved, \beq
\dot{\zeta}_i=0. \eeq

\section{Initial Conditions}

In the previous section we derived a set of equations describing
the long wavelength evolution of fully non-linear gauge-invariant
variables. These equations are exact, as long as the long
wavelength approximation we employed is valid, and are free of any
slow roll assumptions. Of course, one still needs a recipe for
providing initial conditions for the $\cQ^A_i$ variables. We
propose to use linear theory for calculating their values after
horizon crossing. Any further non-linearity is introduced via the
non-linear evolution. We comment below on whether this is good
enough, at least for the cases where interesting effects may
arise.

When linearized, the spatial vectors $\cQ^A_i$ are simply the
gradients of the well known Sasaki-Mukhanov variables. These are
the proper fields to be quantized during inflation \cite{mc,mfb}
and we will assume that they are the appropriate variables for
setting up initial conditions. One way of proceeding would be to
just take \beq \cQ^A_i(t_{in},\mathbf{x})=\partial_i\delta q^A
=\int\limits_{\cV} {d^3k \over (2\pi)^{3\over 2}}{1\over
\sqrt{2}}\left[\cQ^A{}_B(t_{in},k)\,\alpha^B(\mathbf{k})\,ik_i\,e^{i\mathbf{k}\mathbf{x}}+{\rm
c.c.}\right]\,, \eeq where ${\rm c.c.}$ stands for the complex
conjugate. We have taken a finite region in $k$-space, denoted by
$\cV$, which includes the scales of interest and we have defined n
constant complex stochastic quantities $\alpha^A$ with $n$ being
the number of scalar fields, $A=1\ldots n$. The classical random
field exhibits the same correlations as the quantum field if the
stochastic constants $\alpha^A$ satisfy \beq \label{correlation}
\langle\alpha^A\left(\mathbf{k}\right)\alpha_B^{*}\left(\mathbf{k}'\right)\rangle=\delta^{A}{}_B\,\delta\left(\mathbf{k}-\mathbf{k'}\right)\,,\quad
\langle\alpha^A\left(\mathbf{k}\right)\alpha_{B}\left(\mathbf{k}'\right)\rangle=0
\eeq where $\langle...\rangle$ denotes an ensemble average. We
have of course implicitly assumed that on long wavelengths quantum
fields can be considered as classical stochastic fields with the
same correlation properties \cite{classicality, stochastic, sb2}.
The matrix $\cQ^A{}_B(\mathbf{k},t)$ is the solution to the
\emph{linear} equation of motion \cite{bartjan}
\beq\label{lin_eq_matrix} {\cD^2_t}Q^A{}_B-\left({\dot{N} \over
N}-NH\right){\cD_t}Q^A{}_B+\left(\left({Nk \over
a}\right)^2\delta^A{}_C+\Omega^A{}_C\right)Q^C{}_B=0. \eeq The
initial conditions for $\cQ^A{}_B$ are in turn set up when the
relevant mode is deep inside the horizon $\left((k/a)\gg H\right)$
using the properly normalized WKB solutions of
($\ref{lin_eq_matrix}$) in that regime \beq\label{init}
Q^A{}_B(k,t)={1\over
\sqrt{2k}}\exp\left[-ik\int\limits^t_{t_0}{Ndt'\over
a(t')}\right]\delta^A{}_B. \eeq

A more dynamic way of setting up the $\cQ^A_i$ after horizon
crossing involves using source terms on the r.h.s. of
(\ref{basic}) which continuously update the values of $\cQ^A_i$ as
more modes enter the long wavelength system. In this sense these
sources also set up "initial conditions", albeit in a continuous
manner. A heuristic argument for the derivation of such terms,
which is exact at the linear level, is as follows: Start from
linear theory and the full equation of motion for $\delta q^A$
\cite{bartjan} \beq \label{lin_eq} {\cD^2_t}\delta
q^A{}-\left({\dot{N} \over
N}-NH\right){\cD_t}\delta_q^A+\left(\left({Nk \over
a}\right)^2\delta^A{}_C+\Omega^A{}_C\right)\delta q^C{}=0. \eeq
Since we are interested in the long-wavelength semi-classical
dynamics, we can coarse-grain $\delta q^A$ thus defining a long
wavelength linear field \beq \delta\bar{q}^A(\mathbf{x})=\int
d^3x' \delta q^A(\mathbf{x}')W\left(|\mathbf{x}-\mathbf{x}'|\over
R \right), \eeq where $R$ is an appropriate smoothing scale and
the window function $W$ is normalised to unity \beq
(2\pi)^{-3/2}\int d^3x'W\left(|\mathbf{x}-\mathbf{x}'|\over R
\right)=1. \eeq From the convolution theorem the Fourier transform
of $\delta\bar{q}^A(\mathbf{x})$ reads \beq
\delta\bar{q}^A(\mathbf{k})= \delta{q}^A(\mathbf{k})\cW(k), \eeq
where $\cW(k)$ is the Fourier transform of the window function
$W$. From ($\ref{lin_eq}$) we can now derive an equation of motion
for $\delta\bar{q}^A(\mathbf{k})$. It is easily seen to be \beq
\label{lin_eq_long} {\cD^2_t}\delta
\bar{q}^A(\mathbf{k})-\left({\dot{N} \over
N}-NH\right){\cD_t}\delta \bar{q}^A(\mathbf{k})+\Omega^A{}_B\delta
\bar{q}^B(\mathbf{k})=\xi^A(\mathbf{k}), \eeq
 with
\beq\label{lin_noise} \xi^A(\mathbf{k})=\delta q^A(\mathbf{k})
\ddot{\cW}(k) +\left[2D_t\delta
q^A(\mathbf{k})-\left({\dot{N}\over
    N}-NH\right)\delta
  q^A(\mathbf{k})\right]\dot{\cW}(k)-\left(Nk\over a\right)^2\cW(k)
\delta q^A(\mathbf{k}). \eeq In ($\ref{lin_noise}$), $\delta
q^A(\mathbf{k})$ refers to the solution of the full linear
equation ($\ref{lin_eq}$) with subhorizon initial conditions
($\ref{init}$). The real space version of ($\ref{lin_eq_long}$)
reads \beq\label{lin_RS} {\cD^2_t}\delta
\bar{q}^A(\mathbf{x})-\left({\dot{N} \over
  N}-NH\right){\cD_t}\delta \bar{q}^A(\mathbf{x})+\Omega^A{}_B\delta
\bar{q}^B(\mathbf{x})=\dk\xi^A(\mathbf{k})\e+{\rm c.c.}\,. \eeq
The gradient of the l.h.s of ($\ref{lin_RS}$) is the linear
version of ($\ref{basic}$). We will therefore postulate that a
suitable equation for the study of the long wavelength non-linear
dynamics of the system with initial conditions provided from
subhorizon scales by linear theory is \beq\label{main_stoch}
{\cD^2_t}\cQ^A_i-\left({\dot{N} \over
  N}-NH\right){\cD_t}\cQ^A_i+\Omega^A{}_B\cQ^B_i=\dk\,ik_i\,\xi^A(\mathbf{k})\e+{\rm c.c.}\,,
\eeq where now all coefficients depend on the full non-linear
$Q^A_i$. Since $\xi^A$ is stochastic, equation
($\ref{main_stoch}$) is a non-linear Langevin equation for the
long wavelength system.

Equation ($\ref{main_stoch}$) is the main result of this section.
We can use it along with ($\ref{constraints_1}$) -
($\ref{constraints_2}$) to generate non-linear simulations of
inflationary perturbations. Note that the choice of time
(\ref{time}) has one more advantage when used in conjunction with
stochastic noise. The scale that separates the long-wavelength
from the short-wavelength regime for linear perturbations is the
comoving Hubble radius. For this reason we will choose our
smoothing scale $R$ to be a multiple of $(aH)^{-1}$: \beq
R={c\over aH}, \eeq with $c > 1$ \cite{stochastic}\footnote{ The
values $c\sim 3-5$ will do for safely neglecting the $k^2$ term in
$\xi^A$.}. For the gauge choice ($\ref{time}$) the split between
long and short wavelengths is homogeneous throughout all of space
since $aH$ is homogeneous. For other choices of time parameter,
the `time' when a mode is added to the long wavelength system
differs from point to point. It seems conceptually more appealing
to have the noise added simultaneously everywhere and that is
exactly what a $\partial_i\ln (aH)=0$ gauge does. The same
conclusion for slightly different reasons was reached in
\cite{sb2}.

The use of linear theory for setting up initial conditions for the
$\cQ^A{}_i$ or constructing the noise term completely ignores
non-linearities from sub-horizon scales for these variables. One
might question whether such an approximation misses out important
effects before and around horizon crossing. Although this question
cannot be answered within the framework described above, explicit
calculations using this methodology \cite{RSvT05, RSvT05(2)} prove
that such effects are actually unimportant. In the case of both
single and multiple-field inflation perturbative calculations at
tree level \cite{tree} (see also \cite{weinberg} for a
consideration of loop effects) show that non-Gaussianity at the
time of horizon crossing is suppressed by slow-roll factors. This
is considered too low to be observable in any future experiment.
This result is actually reproduced in a simple way using the
methods outlined above \cite{RSvT05, RSvT05(2)}. However, as shown
in \cite{RSvT05(2)} subsequent non-linear evolution in certain
models can enhance non-Gaussianity, making initial non-linearities
subdominant. If this non-Gaussianity is observable, the scheme
proposed above can safely be used to calculate it.

\section{Discussion and future prospects}

The system of stochastic inflation equations presented here
provides a concrete and self-consistent realization for simulating
the generation and evolution of perturbations in the long
wavelength approximation. The method can be applied to arbitrary
single or multi-field inflationary models producing both adiabatic
and isocurvature fluctuations and does not depend on the slow-roll
approximation.  Moreover, this approach incorporates the
nonlinearities inherent in the Einstein equations from the point
the perturbations leave the cosmological horizon until the end of
inflation. The method is, therefore, very relevant to the study of
non-Gaussianity in inflation, which may prove to be a key litmus
test of specific realistic models \cite{NG}. Indeed, the limits on
non-Gaussianity are set to improve substantially over the next few
years with forthcoming CMB experiments and so this may prove to be
an important confrontation with observation. An extension of the
ideas presented here as well as analytic and numerical approaches
for calculating this non-Gaussianity will appear in forthcoming
publications \cite{RSvT05, RSvT05(2)}.

The aim is to estimate the amount of non-Gaussianity generated and
check whether it can be used as a further discriminant between
models. Since this method can provide the real space non-Gaussian
fluctuations, the possibility opens up for applying non-Gaussian
tests both in real and Fourier space and determining which one is
the optimal given a particular model. Such simulations therefore
would be an important advance towards making quantitative
predictions for realistic inflation models which can be tested
against observational data, notably forthcoming CMB experiments.

Here, we shall only briefly describe the numerical implementation
of these methods, leaving a detailed discussion for a much longer
paper \cite{LRSvT05}.  The development of a large and complex
stochastic inflation code is well-advanced but is still undergoing
rigorous testing in specific cases on a parallel supercomputer
(COSMOS). There are basically five key stages to the
implementation: (i) homogeneous (background) solution, (ii) linear
perturbation evolution, (iii) stochastic inflation generation and
initial nonlinear evolution, (iv) subsequent long wavelength or
separate universe evolution and, finally, (v) the end of inflation
and a determination of the resulting (nongaussian) adiabatic and
isocurvature fluctuations, corresponding to the initial conditions
for standard large-scale structure and CMB analysis.  Both (i) and
(iv) are straightforward and essentially identical, solving
(\ref{friedmann}) and (\ref{momentum_dynamics}), although the
separate universe evolution is for a large grid of $N^3$ universes
with perturbed initial conditions and so it is computationally
intensive. Step (ii) entails the solution of the linear mode
matrix equations (\ref{lin_eq_matrix}) for a general single or
multi-field inflation model. The next stage (iii) is the most
complex and computationally intensive when we solve for the ${\cal
Q}_i$'s in the key nonlinear perturbation equations (\ref{basic})
while adding stochastic noise from the linear perturbation
evolution as in (\ref{main_stoch}).  At each timestep in this
evolution, this entails the iterative correction of the separate
universe variables to incorporate the new stochastic fluctuations
in the ${\cal Q}_i$'s.  There are several constraint equations
which are monitored to ensure that this occurs self-consistently
and that the evolution does not drift away from the correct
nonlinear solution. This expensive evolution is only undertaken
while the window function for the stochastic evolution is such
that significant new fluctuations are being added to the numerical
grid, before switching to the more efficient separate universe
evolution (iv). Finally, at the end of inflation (v) we only
implement fairly rudimentary `instantaneous' reheating to obtain a
set of gauge-invariant perturbations suitable for subsequent
perturbation evolution. We are then using the final data products
of nongaussian adiabatic and isocurvature perturbations as input
in a full Boltzmann evolution code to create full-sky and high
resolution CMB realizations for analysis and comparison with
observation \cite{LRSvT05}.

We finish by emphasising the key advantages of this new approach
to the study of nonlinear inflationary fluctuations, while also
pointing out areas for further development of these ideas.  First,
although the discussion of stochastic inflation and the separate
universe approach now has a substantial history
\cite{stochastic,sb2}, in most previous work it has usually been
applied to the field perturbation sector without self-consistently
solving the constraints for the metric perturbations or, in
exceptional cases, that has been achieved only for very specific
single-field models \cite{sb2}.  Here, the method has been applied
to both field and metric perturbations in generic inflation models
with all the constraints satisfied  in the long wavelength
approximation.  Secondly, `generic' includes multi-field models
which are believed to be much more likely to produce significant
non-Gaussianity; indeed, most recent realistic inflation
model-building entails several fields.  Thirdly, the methods
presented here allow for the nonlinear evolution of the
perturbations from the time at which each mode leaves the horizon.
This takes into account the effect of the long wavelength modes on
the subsequent shorter wavelength noise.  Fourthly, unlike
previous work, this long wavelength framework allows for general
choices of time-slicing with the relevant perturbation variables
always remaining invariant. Finally, the separate universe
approach considered here corresponds to the lowest order terms of
a gradient expansion of the full nonlinear Hamilton-Jacobi
equations \cite{SalSte95}. This seems to the authors likely to be
a more fruitful and elegant approach to incorporating further
nonlinearity than the technically much more complicated
alternative of applying higher order perturbation theory to the
original Einstein equations \cite{secondorder}.  Of course, other
improvements can be made to the methods discussed here, such as
including the subdominant effects of vector and tensor modes and a
more rigorous justification of the quantum noise. Nevertheless, we
believe the present work constitutes an important step forward
since it enables the quantitative calculation of nonlinear effects
in generic inflationary models.

\section*{Acknowledgements}

\noindent We are most grateful to Bartjan van Tent for his
critical input and a fruitful collaboration on clarifying,
extending and applying this formalism. The results of this
collaboration are reported elsewhere \cite{RSvT05, RSvT05(2)}.
Misao Sasaki and Niayesh Afshordi pointed out an unnecessary
assumption in the first version of this work which we removed in
this version. This research was supported by PPARC grant
PP/C501676/1.

\appendix

\section{Derivation of Equation ($\ref{basic}$)}

In this appendix we outline the derivation of equations
($\ref{basic}$). Taking the covariant time derivative of $\cQ^A_i$
we get \beq
\cD_t\cQ^A_i=NH\cQ^A_i+a\cD_i\left(N\Pi^A\right)-a\partial_i\left(NH\right){\Pi^A
  \over H}-aX_i\cD_t\left({\Pi^A\over H}\right),
\eeq where we have used that
$\cD_t\left(\partial_i\phi^A\right)=\cD_i\left(N\Pi^A\right)$. We
then take the second time derivative. Most of the calculation is
straightforward and we will not reproduce it here since it leads
to long expressions. We only discuss non trivial points. In
calculating $\cD^2_t\cQ^A_i$ we encounter the term
$\cD_t\cD_i\Pi^A$. Using the definitions (\ref{cov_1}) we readily
see that \bea
 \cD_t\cD_i\Pi^A&=&\cD_i\cD_t\Pi^A+NR^A{}_{FCB}\Pi^F\Pi^C\partial_i\phi^B\nonumber\\
&=&\cD_i\cD_t\Pi^A+{N\over a}R^A{}_{FCB}\Pi^F\Pi^C\cQ^B_i. \eea
Here, $R^A{}_{FCB}$ is the curvature tensor of the field manifold
and in deriving the last equality we used the antisymmetry
properties of its last two indices. Now we can readily evaluate
the $\cD_i\cD_t\Pi^A$ term by taking a covariant spatial
derivative of ($\ref{momentum_dynamics}$) so we have \beq
\cD_t\cD_i\Pi^A =
-3NH\cD_i\Pi^A-3N\Pi^A\partial_iH-3\partial_iNH\Pi^A-\partial_iNV^A-NV^A{}_B\partial_i\phi^B+{N\over
a}R^A{}_{FCB}\Pi^F\Pi^C\cQ^B_i. \eeq We can now replace
$\cD_i\Pi^A$ wherever it occurs by \beq \cD_i\Pi^A={1\over
N}\cD_t\left[\left({1\over
    a}\cQ^A_i\right)+X_i{\Pi^A\over H}\right]-{1\over N}\Pi^A\partial_iN,
\eeq and substitute $\partial_i H$ from \beq
\partial_i H=-{4\pi \over m_{\rm pl}^2}\Pi_B\left({1\over
    a}\cQ^B_i+{\Pi^B\over H}X_i\right).
\eeq Another term that needs to be dealt with is
$\partial_i\dot{H}$. By differentiating ($\ref{H_dynamics}$) we
get \beq
\partial_i\dot{H}=-\grv\partial_iN\Pi^2-{8\pi\over m_{\rm
    pl}^2}N\Pi_B\cD_i\Pi^B.
\eeq The second term on the r.h.s can be obtained from the
differentiation of ($\ref{friedmann}$) along with the use of
($\ref{0i}$). It reads \beq \Pi_B\cD_i\Pi^B=-{1\over a}
\left(3H\Pi_C+V_C\right)\cQ^C_i-\left(3\Pi^2+{V_C\Pi^C\over
  H}\right)X_i.
\eeq Putting everything together and using eq. ($\ref{Xi}$) we see
that the terms involving $X_i$ and $\partial_iN$ cancel out and we
are left with eq ($\ref{basic}$).

\end{document}